\documentclass[aps, twocolumn]{revtex4-1}

\usepackage{graphicx}
\usepackage{amsmath}
\usepackage{dsfont}
\usepackage{amssymb}
\usepackage{physics}
\usepackage{hyperref}
\usepackage{ulem}
\hypersetup{colorlinks=true,
	    final=true,
	    linkcolor=blue,
	    citecolor=blue,
	    filecolor=blue,
	    urlcolor=blue,}

\usepackage{amssymb}

\usepackage{wrapfig}

\newcommand{\pr}{\partial}
\newcommand{\rta}{\rightarrow}

\newcommand{\ep}{\epsilon}

\newcommand{\om}{\omega}
\newcommand{\ra}{\rangle}
\newcommand{\la}{\langle}

\newcommand{\beq}{\begin{equation}}
\newcommand{\eeq}{\end{equation}}

\begin{document}

\title{One Rudolf Peierls' surprise: the quantum-to-classical transition in the context of solid-state physics}

\author{Navinder Singh}

\email{navinder.phy@gmail.com; Phone: +91 9662680605}
\affiliation{Theoretical Physics Division, Physical Research Laboratory (PRL), Ahmedabad, India. PIN: 380009.}

\begin{abstract}
In solid state physics, it is an unsaid (tacit) assumption that the Bloch theorem is applicable to a crystal lattice even if it is of the macroscopic dimensions, provided periodicity is maintained. However, in a realistic situation, electrons in a periodic potential of ions constitute an open quantum system and are subjected to decoherence and dissipation. A natural question arises: up to what distances electrons in a periodic potential can be considered as constituting an effective closed quantum system? And what is the cause of decoherence? To answer some of these questions, the seminal theory of Ovchinnikov and Erikhman of decoherence due to ionic motion is revisited and an oversight of the authors is corrected. Correct conditions for decoherence to occur are worked out. Length scale up to which the motion of ions remains coherent is also calculated. Finally, a realistic physical picture is discussed. 
\end{abstract}

\maketitle

 \section{Rudolf Peierls' surprise}
 
Rudolf Peierls, in section 3.1, in his famous book entitled ``Surprises in theoretical physics"\cite{peierlsb}, argues that the application of the Pauli exclusion principle to a macroscopic metallic sample is no less than a surprise! Once the Pauli exclusion principle and hence the Fermi-Dirac statistics is applied to all the conduction electrons, the problems of the classical statistical mechanics of electrons in metals immediately get rectified. There are several examples of it, and weak Pauli paramagnetism of metals is one of them.  At zero temperature, due to the presence of two electrons of opposite spin in each momentum state (up to the maximum Fermi momentum) spins are all paired up and not allowed to point in the direction of the applied magnetic field. At a finite temperature, only a very small fraction of electrons ($\sim\frac{k_BT}{E_F}$) in the diffusion zone of width $k_BT$ around the Fermi surface are allowed to point in the direction of the applied field  and can show paramagnetism. The product of the width of the diffusion zone $\frac{k_BT}{E_F}$ (proportional to the number of available electrons) with the Curie susceptibility of each electron which is $\propto\frac{1}{T}$ leads to a net susceptibility independent of temperature, as first shown by Pauli.  The application of Pauli principle and hence Fermi-Dirac statistics to the conduction electrons solved other problems, such as that of the heat capacity etc\cite{am,navb}.

 According to Rudolf Peierls, the surprise lies in the following observation. Peierls argues\cite{peierlsb}:
 
 \begin{quote}
{\texttt{``If two electrons are at the opposite ends of a metal wire of macroscopic dimensions, say a meter length, is it not surprising that they can manage to avoid being in the same state of motion? How can each of them know what the other is doing?''   --- Rudolf Peierls. } }
\end{quote} 
  
 The question is:  up to what distance the location of an electron can be specified such that quantum-statistical effects remain intact? (of course, this must be done within the leeway provided by the Heisenberg uncertainty principle). Is there a measurement process going-on of the electrons position coordinates by the lattice degrees-of-freedom irrespective to the fact that whether there is a conscious observer present or not?  Do electrons at the right hand side of the wire constitute a distinct Fermi-Dirac distribution than those towards the left end of the wire (but with the same chemical potential)? Or, are the electrons just running waves with position uncertainty of the order of the length of the wire (one meter in this case), thus forming a single Fermi-Dirac distribution?  In other words, if an electron's location within some small length interval towards the right hand side of the wire is specified (by some method), then this state of the electron will be orthogonal to the state of another electron whose location is specified in a small length interval (much less than the length of the wire) towards the right hand side of the wire. Then these two states must be counted (for the Pauli principle) as distinct states of motion.

 To delve deeper into the issue, Peierls introduces a simplified picture. Consider 1-D "box" of length $L$.  With periodic boundary conditions.  The $k-$space density of levels is given by $\frac{dn}{dk} = \frac{L}{2\pi}$. The electrons are running waves in the entire lattice of length $L$. The right hand side or the left hand side division of electrons is not possible now. To mimic a realistic situation, however, Peierls introduces electron wave-packets  of finite extension (instead of running waves in the entire lattice). Let us say that $a$ is the "lattice constant" of the "box" ($a<<L$). Let these wave-packets are localized on a length scale of the order of $a$ (figure 1). Each "cell" of length $a$ can have many wave-packet states (figure 1).

 \begin{figure}[h!]
    \centering
    \includegraphics[width=1.0\columnwidth]{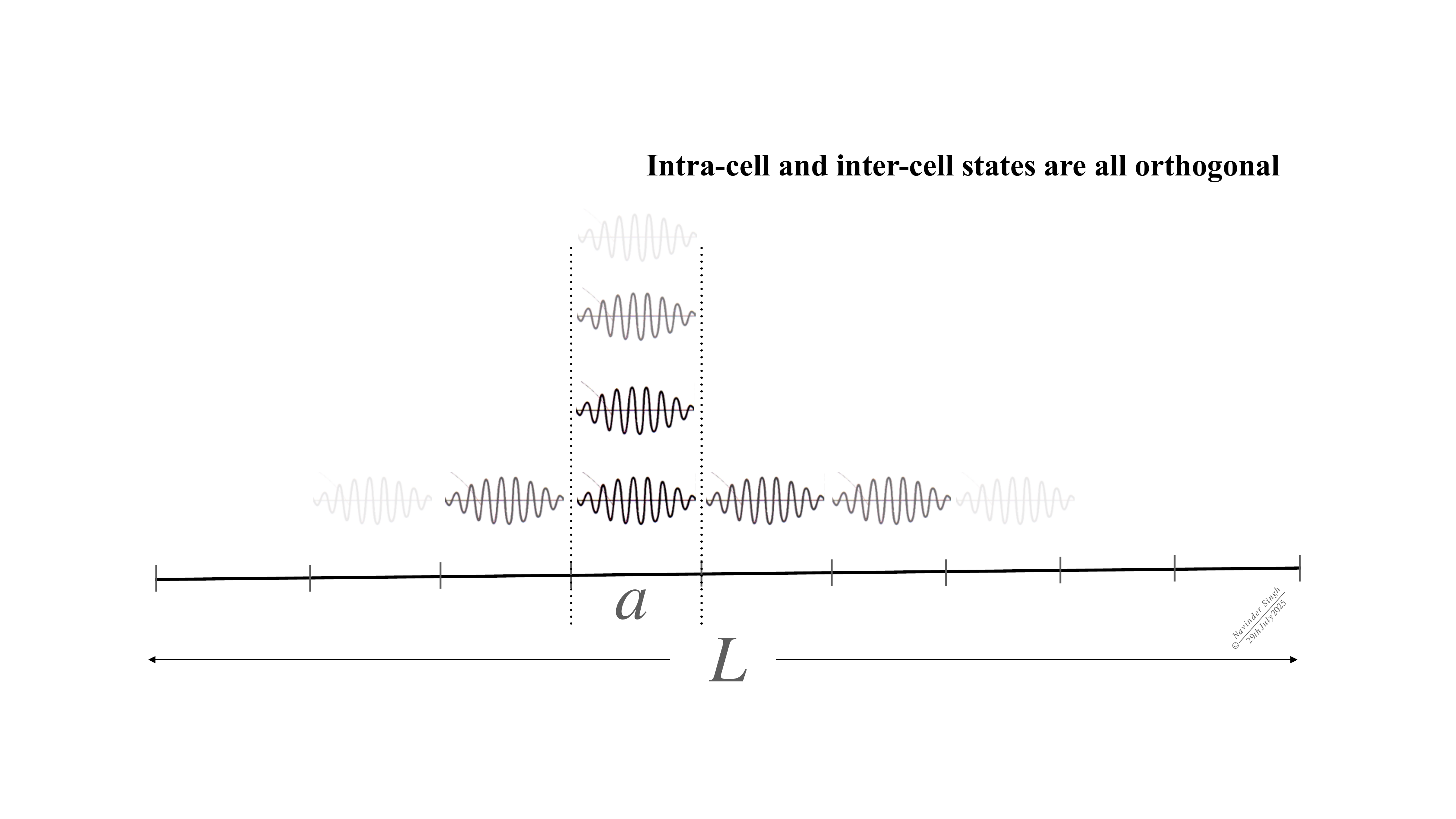}
    \caption{Electrons as wave-packets for counting states in different cells. Each cell will have many wave-packet states.}
  \end{figure}

Now the $k$-space density is reduced to $\frac{a}{2\pi}$ (by a factor of $a/L$) as compared to first case. It turns out that states in different cells and states in each cell are all orthogonal states and must be counted distinct in the Pauli principle. {\it The electron wave-packets in a cell towards the left hand side has a Fermi-Dirac distribution independently of the ones towards the right hand side of the cell. }
 
A rough estimate of the size of the electron wave packets can be given as follows. Energy uncertainty at temperature $T$ is of the order of $k_B T$. The momentum uncertainty, and hence, $\Delta k = \frac{mk_BT}{\hbar^2k_F}$, which corresponds to $\Delta x \sim  100~nm$ for a metal at room temperature with Fermi energy $1~eV$. Thus, the electronic states separated over 100 nanometer length scales must be counted as separate states of motion.

To see that there is a problem with the picture of electrons as running waves throughout the lattice (of macroscopic dimensions), consider again the case of 1-D lattice of $N$ sites with lattice constant $a$. Assume that at each lattice site there is a Wannier  state that we represent with $|1\ra,~|2\ra,~|3\ra,.......|N\ra$. The state of an electron in the lattice is represented by the superposition state over all the possible states:

\beq
|\phi\ra = \sum_{n=1}^N c_n|n\ra.
\eeq

The point which is most important is that the above equation is true for an isolated quantum system. Now consider a realistic situation. At each site there is an ion sitting which is oscillating about its equilibrium position both due to zero point quantum motion and due to thermal vibrations. {\it The question is: Whether the electron in this 1-D lattice with underlying thermally vibrating lattice system of sites is still an isolated quantum system? Answer to this question is no. We are now dealing with an open quantum system.} In addition, ions do not completely randomly vibrate from each other. They are connected with each other by some sort of bonds and their vibrations about their equilibrium positions correlate with each other to give collective oscillations (phonon modes). 

If it is an open quantum system then how does the Fermi-Dirac distribution of electrons in a metal remain intact? Its presence and stability is shown by so many experimental facts (specific heat capacity of metals, Pauli paramagnetism etc). Thus the problem seems formidable at the outset. It turns out that over a time interval (to be specified in subsequent sections), the motions of nearby ions become un-correlated. This brings into the problem certain kind of a random component. If the tunneling matrix element  (the overlap integral) between two near neighbor ions is $-\alpha$, then because of the random component in the vibrations of the ions, there exits a small random overlap integral ($\beta(t)$). This is because the overlap integral depends on the bond length and there is a random component in the oscillations of the ions from their equilibrium positions, thus in the bond length. The total overlap integral now is given by $-\alpha -\beta(t)$; such that:

\beq
\overline{\beta(t)} =0,~~~~~\overline{\beta(t)\beta(t')} = \beta_0\delta(t-t').
\eeq

That is, the random component of the overlap integral has zero mean and it is delta correlated. This model of noise was used in the pioneering investigation on the motion of a quantum particle in a stochastic medium by A. A. Ovchinnikov and N. S. Erikhman\cite{oe}. The electronic component in a metal is in fact an open quantum system subjected to dissipation and decoherence, and one can understand why the picture of electrons as running waves throughout the lattice cannot be true. 

Armed with this understanding, it is possible to quantitatively answer the problem posed by Peierls' surprise, and Ovchinnikov-Erikhman (OE) program provides a scheme for it. However, we will point out a drawback (or an oversight) in the analysis of Ovchinnikov and Erikhman, and show that the random component occurs over a time scale (of the order of pico-seconds) much larger than the time scale over which electrons move (of the order of femto-seconds) and superposition principle remains valid to a larger (much greater than the lattice constant) length scale (but surely not throughout the lattice) dictated by the position correlator which we compute. Therefore, the present work aims to put the electron wave-packet picture on a quantitative footing.

This paper is organized in the following way: In section II we review the Ovchinnikov-Erikhman program of decoherence. In section III we will point out a serious problem with it. In section IV we introduce a realistic picture in which we will calculate the position-position correlator to investigate up to what length scale  the motion of ions remain correlated and on what length scale the random component begins to emerge. Section V includes a thorough discussion and we end the manuscript after concluding in section VI.

\section{The Ovchinnikov-Erikhman (OE) program}

Ovchinnikov and Erikhman\cite{oe} consider a 1-D lattice with the Hamiltonian:

\begin{eqnarray}
H &=& \frac{1}{2}\sum_{m,n=-\infty}^{+\infty}\beta_{mn}^0 (|m\ra\la n|+|n\ra\la m|)\nonumber\\
&+&\frac{1}{2}\sum_{m,n=-\infty}^{+\infty}\beta_{mn}(t) (|m\ra\la n|+|n\ra\la m|).
\end{eqnarray}

Here $\beta_{mn}^0$  constitute regular tunneling matrix elements for $m \ne n$, and on-site energies for $m=n$, whereas $\beta_{mn}(t)$ is random tunneling matrix element which describes the action of the stochastic medium on the system.  To appreciate the point which we want to make clear here, a two-level system in a stochastic environment is  also sufficient. This is considered by OE as a warm-up exercise in their paper. The two-level system has the following Hamiltonian:

\beq
H = 0|1\ra\la1| +\ep |2\ra\la2| - (\alpha +\beta(t)) [|1\ra\la2|+|2\ra\la1|],
\eeq
where we neglect stochasticity at the site energies ($\ep$ is constant), and only the tunneling matrix element (the overlap integral) has a random component $\beta(t)$ with the properties given by equation (2). The equation of motion (EOM) for the density matrix is given by:

\beq
i\frac{d\rho}{dt} = [H,\rho].
\eeq
Here $\hbar=1$. These can be written in the site basis\cite{oe}. One needs to find the averaged density matrix (averaged over the stochastic noise). The following set of equations can be easily derived:

\begin{eqnarray}
i \frac{d\bar{\rho}_{11}}{dt} &=& \alpha (\bar{\rho}_{12} -\bar{\rho}_{21}) +\overline{\beta(t)\rho(t)_{12}} - \overline{\beta(t)\rho(t)_{21}}\nonumber\\
i \frac{d\bar{\rho}_{22}}{dt} &=& \alpha (\bar{\rho}_{21} -\bar{\rho}_{12}) +\overline{\beta(t)\rho(t)_{21}} - \overline{\beta(t)\rho(t)_{12}}\nonumber\\
i \frac{d\bar{\rho}_{12}}{dt} &=& -\ep \bar{\rho}_{12} +\alpha( \bar{\rho}(t)_{11} - \bar{\rho}(t)_{22}.
\end{eqnarray}

In the averaged equations, products of the form $\overline{\beta(t)\rho(t)_{12}}$ are encountered which can be decoupled by using what is called the Novikov theorem\cite{oe}. If we write $\bar{\rho}_{11} = \frac{1}{2}+\rho,~\bar{\rho}_{22} = \frac{1}{2}-\rho,~\bar{\rho}_{12} = \rho_1 +i\rho_2$, and $\rho_{12}$ is a complex conjugate of $\rho_{21}$, then the EOM takes the form:

\begin{eqnarray}
\dot{\rho} &=& 2\alpha \rho_2 - 4 \beta_0 \rho \nonumber\\
\dot{\rho}_1 &=& -\ep \rho_2\nonumber\\
\dot{\rho}_2 &=& \ep \rho_1 - 2 \alpha \rho - 4\beta_0 \rho_2,
\end{eqnarray}
which can be solved easily to give, for example, the time dependence of the off-diagonal element as:

\beq
\rho_1(t) = e^{-2\beta_0 t} \left(A \cosh(\om t) + \frac{\ep +\beta_0 A-2\alpha B}{\om}\sinh(\om t)\right).
\eeq
Here $\om = \sqrt{\ep^2 +4\alpha^2 - 4\alpha\beta_0}$, and $A$ and $B$ are the values of the off-diagonal elements at time $t=0$. With diagonal elements at time $t=0$ set as: $\rho_{11} =1,~~\rho_{22} =0$. {\it The important point to be noted is that the off-diagonal elements of the density matrix decay on a time scale set by $\frac{1}{\beta_0}$ which is the coefficient of the delta correlated stochastic perturbation (equation 2). The diagonal elements reach the value $\frac{1}{2}$ on the  same time scale. It means that the initial state which can be written as a quantum mechanical superposition state over $|1\ra$ and $2$ as $a|1\ra+b|2\ra$, decoheres,  and now that superposition is destroyed on a time scale of the order of $\frac{1}{\beta_0}$. This is a simplest version of the quantum-to-classical transition. The position states $|1\ra$ and $|2\ra$ are the pointer states\cite{zurek1}}.

OE argues that the stochastic component originates from the underlying lattice vibrations. They show that in the high temperature limit $k_B T>>\Theta_D$ (where $\Theta_D$ is the Debye temperature) positions of the two nearby ions start to exhibit random behaviour, and the properties (given by equation (2)) can be derived.

However, in the following section, we will point out that the delta correlated stochastic noise is applicable {\it if and only if} the time scale involved is much greater than the inverse of the Debye frequency ($t>>\frac{1}{\om_D}$). This time scale turns out to be in pico-seconds for typical metals whereas electron motions takes place on femto-second time scales. {\it Thus the foundation on which the OE program rests must be revised, and there is much more to it.}

\section{Trouble with Ovchinnikov-Erikhman program}

Consider a chain of vibrating atoms and define $x(t)$ as distance between any two nearby atoms at time $t$ (figure (2)).
\begin{figure}[h!]
    \centering
    \includegraphics[width=1.0\columnwidth]{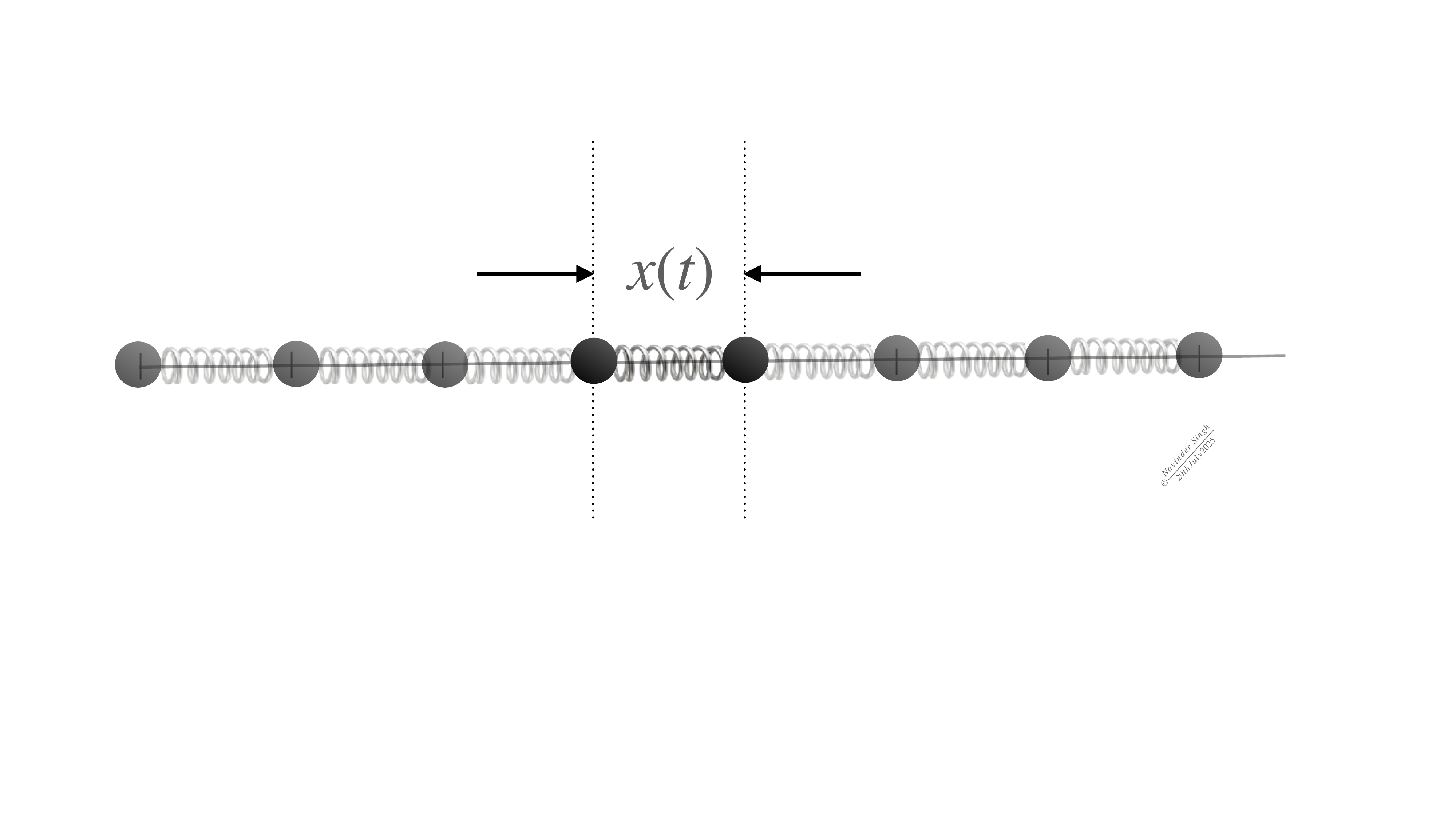}
    \caption{Distance  $x(t)$ between any two nearby atoms as a function of time.}
    \label{f1}
  \end{figure}
Consider longitudinal acoustic vibrations such that the wave vector $q$ is along the lattice constant vector $a$ joining the two atoms. The distance between any two nearby atoms $x(t)$ can be written in terms of the normal modes of vibrations\cite{oe,blochbook}:
\beq
x(t) = \frac{1}{N^{3/2}}\sum_q A_q\cos(\om_q t +\phi_q)\cos(qa).
\eeq
Here $\om_q,~A_q,~\phi_q$ are the frequency, amplitude, and phase of the normal mode defined by wave vector $q$, respectively. $N$ is the number of unit cells with volume of each unit cell $a^3$. $A_q$ and $\phi_q$  are taken to be random quantities for different modes. Consider the following correlator at two different times $t_1$ and $t_2$:

\begin{eqnarray}
\la x(t_1)x(t_2)\ra &=& \frac{1}{N^3}\sum_{q,q'}\la A_q A_{q'}\ra \cos(qa)\cos(q'a)\nonumber\\
&\times&\la\cos(\om_q t_1+\phi_q)\cos(\om_{q'}t_2 + \phi_{q'})\ra.
\end{eqnarray}

Random quantities with different $q$ are independent from each other, and thus independent averages of each vanish. Only the terms which survive are those where $q = q'$,  such that $\la A_q A_{q'}\ra = \la A_q^2\ra$ etc.  The correlator, then, is equal to:
\beq
\frac{1}{N^3}\sum_q\la A_q^2\ra\cos[\om_q(t_1-t_2)]\cos^2(q a).
\eeq

Under the harmonic approximation, the variance of the amplitude of oscillations is connected with the total energy:

\beq
\la A_q^2\ra = \frac{\hbar}{3 M \om_q} \left(\la n_q\ra +\frac{1}{2}\right).
\eeq
Here $M$ is the mass of the ions and $\la n_q\ra = \frac{1}{e^{\beta\hbar\om_q}-1},~~\beta = \frac{1}{k_B T}$. Converting the summation over $q$ into integration:

\beq
\frac{1}{N^3}\sum_q \rta (unit~cell~volume)\int\frac{d^3q}{(2\pi)^3},
\eeq
and using the Debye model for the acoustic phonons ($\om_q = c_s q$, where $c_s$ is the sound speed), and after some simplifications we get the following expression for the correlator:

\begin{eqnarray}
\la x(t_1)x(t_2)\ra &=& \frac{a^3\hbar}{6\pi^2Mc_s^3}\int_0^{\om_D} d\om \om \left(\frac{1}{e^{\beta\hbar\om}-1}+\frac{1}{2}\right)\nonumber\\
&\times&\cos[\om(t_1-t_2)]\cos^2(\om a/c_s).
\end{eqnarray}
Here $\om_D$ is the upper cut-off (the Debye frequency). On changing variables $x = \beta\hbar \om$, we get

\begin{widetext}
 \beq
\la x(t_1)x(t_2)\ra = (aq_D)^3 \left(\frac{k_BT}{\hbar\om_D}\right)^2\left(\frac{\hbar}{6\pi^2M\om_D}\right)\int_0^{\frac{\hbar\om_D}{k_BT}} dx x\left(\frac{1}{e^x-1}+\frac{1}{2}\right)\cos\left[x \frac{t_1-t_2}{(\hbar/k_BT)}\right]\cos^2\left(x \frac{k_BT}{\hbar\om_D} a q_D\right).
\eeq
\end{widetext}
Here $q_D$ is the Debye wave vector. The first cubed term $(aq_D)^3$  is dimensionless, the second term $ \left(\frac{k_BT}{\hbar\om_D}\right)^2 $ is also dimensionless. The 3rd term $\left(\frac{\hbar}{6\pi^2M\om_D}\right)$ has the dimensions of length squared (as it should). As introduced, $x =\frac{\hbar \om}{k_BT}$ is dimensionless. In the denominator of the argument of cosine we have a time scale $\hbar/k_BT$, let us call it the ``thermal" timescale. 

Now, the overlap integral $(\alpha +\beta(t))$ is proportional (in the leading order of approximation) to the instantaneous value of the bond length $x_0 + x(t)$. In the linear approximation: $\beta(t) = c x(t)$, where $c$ is constant. As depicted in equation (2), $\beta(t)$ is assumed to be delta correlated. Let us ask under what conditions the correlator (depicted in equation (15)) goes over to a delta function in time? So that equations (2) can be derived. The correlations decay in the high temperature limit. Let us take the high temperature limit $k_BT>>\hbar\om_D$, and in this limit $x<<1$, the integrand can be simplified:
\begin{eqnarray}
\la x(t_1)x(t_2)\ra &\simeq& (aq_D)^3 \cos^2(a q_D) \left(\frac{k_BT}{\hbar\om_D}\right)\left(\frac{\hbar}{6\pi^2M\om_D^2}\right)\nonumber\\
&\times&\left(\frac{\sin[\om_D(t_1-t_2)]}{t_1-t_2}\right).
\end{eqnarray}
And under the condition $\om_D (t_1-t_2)>>1$ (that is, when time scales involved are much greater than inverse of the Drude frequency) it reduces to:
\beq
\la x(t_1)x(t_2)\ra \sim Const ~T~ \delta(t_1-t_2),
\eeq
where we used $\lim_{a\rta\infty}\frac{\sin(at)}{t} = \pi\delta(t)$. That is, the delta correlated form of the correlator  is obtained when:

\begin{enumerate}
\item $k_B T >>\hbar\om_D$.
\item $|t_1-t_2|>>\frac{1}{\om_D}$.
\end{enumerate}

\begin{figure}[h!]
    \centering
    \includegraphics[width=1.0\columnwidth]{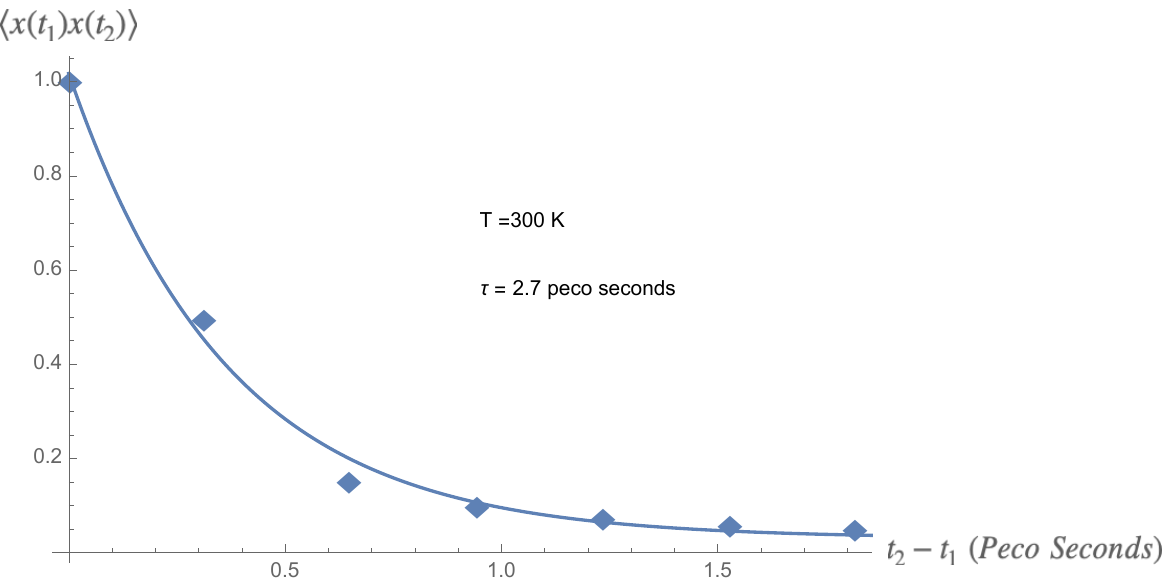}
    \caption{Envelop of the correlator (equation(15)) as a function of time. The calculation is done for metal gold (Debye temperature of gold = 165 $K$, sound speed = 3240 $m/sec$). The data points are the envelop of equation (15) and the solid line is an exponential $e^{t/\tau}$ fit. Least square fitting leads to $\tau = 2.7~psec$.}
  \end{figure}

However, OE argued that the delta correlated form of the correlator $\la x(t_1) x(t_2)\ra$ requires only the high temperature limit\cite{oe}. This is clearly an oversight! The second condition $ |t_1-t_2|>>\frac{1}{\om_D}$  (which is not taken into account by OE) has very important implications that cannot be neglected to reach to a realistic physical picture. To appreciate this point, consider the time scale on which an electron hops from one site to a nearby site. The tunneling matrix element is $\alpha$. The characteristic timescale of this motion is $\tau = \frac{\hbar}{\alpha}$. For $\alpha = 1~eV$, the characteristic timescale $\tau$ is in femtoseconds, whereas the ionic motion occurs on a typical timescale  of $\frac{1}{\om_D}$, which is in picoseconds, much larger than the timescale on which electronic motion take place.  In fact, if the correlator (given in equation (15)) is numerically calculated and plotted (without any approximation), the time scale of its decay turns out to be in picoseconds (refer to figure (3)). Thus the stochastic component emerges when condition (2) [mentioned above] is met.  This also means that the stochastic component does not emerge on the timescale of electronic motion, and electrons can hop from atom to atom without suffering decoherence. But at the same time this coherence cannot be maintained on a macroscopic length scale!

In the following section, we perform another calculation to find out up to what length scale the motion of ions remains coherent, and at what length scale the randomness begins to appear, even when the electronic timescale of femtoseconds are taken into account. This randomness is the actual cause which is behind the Peierls' wave-packet picture of finite extension of the electrons. The incoherence of the lattice degrees-of-freedom ultimately influences and brings the decoherence in the electronic motion.

\section{A realistic picture}

Instead of considering the correlation of the distance between two nearby atoms at different times (as done in section III), consider correlation $\la x_1(t_1)x_n(t_2)\ra$ between the position of the first atom (say) from the origin $x_1(t_1)$ and the position of the $n^{th}$ atom from origin  $x_n(t_2)$ ($n a$ units of length away from origin). In this case equation (10) modifies to:

\begin{eqnarray}
\la x(t_1)x_n(t_2)\ra &=& \frac{1}{N^3}\sum_{q,q'}\la A_q A_{q'}\ra \cos(qa)\cos(n q'a)\nonumber\\
&\times&\la\cos(\om_q t_1+\phi_q)\cos(\om_{q'}t_2 + \phi_{q'})\ra,
\end{eqnarray}
where instead of $\cos(qa)\cos(q'a)$, we now have $\cos(q a)\cos(q' n a)$. Following the similar steps (as done in section III) we get the following equation for the correlator: 
\begin{widetext}
 \beq
\la x(t_1)x_n(t_2)\ra =  \left(\frac{k_BT}{\hbar\om_D}\right)^2\left(\frac{\hbar (aq_D)^3}{6\pi^2M\om_D}\right)\int_0^{\frac{\hbar\om_D}{k_BT}} dx x\left(\frac{1}{e^x-1}+\frac{1}{2}\right)\cos\left[\frac{x(t_1-t_2)}{(\hbar/k_BT)}\right]\cos\left(x \frac{k_BT}{\hbar\om_D} a q_D\right) \cos\left( n x \frac{k_BT}{\hbar\om_D} a q_D\right).
\eeq
\end{widetext}
Which is similar to equation (15), but we now have two cosine factors, one having the site index $n$ in it (the last factor in equation (19)). As argued in the previous section electronic motion is an order of magnitude faster than the ionic motion, we are interested in the position correlation between two far away atoms but at the same instant of time. That is, we are interested in $t_1=t_2$ limit of the correlation function. The above expression takes a simplified form in the high temperature limit $k_BT>>\hbar\om_D$, where we will be able to observe an explicit dependence of the correlation function on the position index $n$. Following the similar high temperature expansion ($x<<1$ limit of equation (19)), we get:

\beq
\la x_1(0)x_n(0)\ra =  \left(\frac{k_BT}{\hbar\om_D}\right)\left(\frac{\hbar (aq_D)^2 \cos(aq_D)}{6\pi^2M\om_D}\right)\left[\frac{\sin(n a q_D)}{n}\right].
\eeq
The envelop of the function and an overall dependence on temperature goes like this:
\beq
\la x_1(0)x_n(0)\ra \sim  T\times \left(\frac{1}{n}\right).
\eeq
This shows a power law decay ($1/n$ decay) of the correlator as a function of $n$. That is, at sufficiently large $n$, the correlation is weak such that it can be neglected. It has non-trivial temperature dependence. The position correlation enhances with temperature. It can be appreciated by using the following argument. The high temperature limit of the equation (12) reads:
\beq
\la A_q^2\ra \simeq \frac{\hbar}{3 M \om_q}\frac{k_B T}{\hbar \om_q}.
\eeq
This means that the variance of the amplitude of oscillations increases with increasing temperature, and correlation magnitude is larger. However, we are having a problem in defining a characteristic length scale in this case. The power law decay with $n$ does not have a characteristic length scale involved in it (unlike in an exponential decay $e^{-l/l_0}$ where it is possible to define a characteristic length scale $l_0$). However, we can ask the following question: after what  value of $n$ the magnitude of the correlation $\frac{x_1(0)x_n(0)}{x_1(0)x_1(0)} =\frac{1}{n}$ becomes negligibly small.  One reasonable value is $n=100$ where the ratio is $0.01$. Thus beyond 100 lattice sites, the correlation becomes negligibly small. If we take the value of the lattice constant as $a=4.1\AA$ (which is the value of the lattice constant for metal gold) we get a "lattice correlation length" of the order of $400\AA=40~nano-meters$.

\section{Discussion}

Above calculations has important implications. In the elementary introduction to solid state physics (refer, for example to\cite{am}), it is a tacit assumption that the Bloch theorem is applicable to a crystal lattice (periodic lattice) even if it is of the macroscopic dimensions (say, for example, a sample of one meter length). But the application of the Bloch theorem to a real metals is not straightforward, even when the lattice is perfect. The question is of decoherence.  As electrons hop from site to site their tunneling is coherent (non-random overlap integral) only up to a characteristic length scale (controlled by the lattice degrees-of-freedom). The length scale is set by the correlator (equation(20)). However, randomness does emerge over a length scale (``the lattice correlation length", as pointed out in the last section) which is much greater than the lattice constant. It is given by the correlator (equation (20)). This must be kept in mind before applying the Bloch theorem. This also provides a concrete answer to the surprise pointed out by Peierls regarding the counting of states (for the Pauli principle) of the far-separated electrons.  

In addition, the standard picture on which the entire semi-classical treatment of electron dynamics is based\cite{am,navb} now secures a concrete basis. It is assumed that electrons are in the form of wave packets of real space extension much greater than the lattice constant, but, at the same time, much smaller than the wavelength of the external applied fields (refer to figure (4)).

However, this picture is introduced ad-hoc\cite{am,navb}. The above calculations (of the lattice correlation length) provide a foundation to this ad-hoc picture. The position uncertainty (of the order of the lattice correlation length) given by the position correlator in equation (20) is much greater than the lattice constant but at the same time can be much smaller than the wave length of the applied fields. These considerations are very important as far as the foundations of the semi-classical theory of electron dynamics are concerned, and provide {\it a-priori} justification to it. 
\begin{figure}[h!]
    \centering
    \includegraphics[width=1.0\columnwidth]{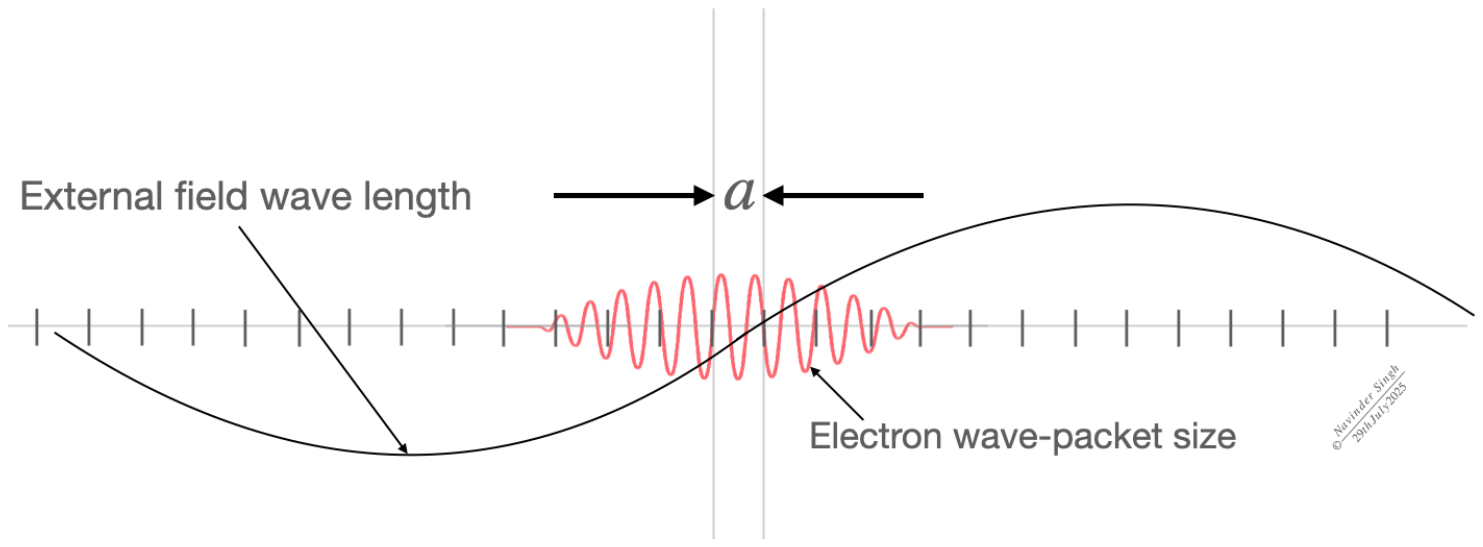}
    \caption{The standard foundational picture of the semi-classical electron dynamics\cite{am,navb}. Electrons are in the form of wave-packets of finite spatial extension, much bigger than the lattice constant. But at the same time, the size of these wave packets is much smaller than the wavelength of the external applied fields.}
  \end{figure}
The problem of electrons hoping from site to site with underlying thermally vibrating lattice can be gauged from a different and a well established point of view\cite{zurek1}.  This point of view originated from the studies\cite{fv,dek,cal1,cal2,joos,paz} of the dynamics of a particle interacting with a thermal bath which consists  of a collection of harmonic oscillators. In the high temperature limit (neglecting zero point fluctuations) the reduced density matrix of the particle in the position representation obeys the following master equation\cite{zurek1}:

\begin{widetext}
 \beq
\frac{d\rho(x,x')}{dt} = \frac{i}{\hbar}[H,\rho] -\gamma  ~(x-x') \left(\frac{\pr}{\pr x} -\frac{\pr}{\pr x'}\right)\rho - \frac{2 m\gamma k_BT}{\hbar^2} (x-x')^2\rho.
\eeq
\end{widetext}
Here $\gamma$ is the relaxation (dissipation) rate. As Zurek argued\cite{zurek1} the last term of the above master equation can be used to estimate the decoherence time scale. It turns out that\cite{zurek1}:

\beq
\tau_D =\tau_R \left( \frac{\hbar}{\Delta x \sqrt{2 mk_BT}}\right)^2.
\eeq
Here $\tau_R =\frac{1}{\gamma}$ is the relaxation time, and $\tau_D$ is the decoherence time (on which the off-diagonal elements of the density matrix decay).

Let us compute the value of $\tau_D$ for our problem of electron hoping from site to site. If we take $\Delta x = a = 4.1\AA$ (value of the lattice constant for metal gold) and $T=300~K$ (room temperature), then we get $\tau_D \simeq 10 \tau_R$.  This is an order of magnitude greater than the relaxation time. The actual value of it can be estimated. The Drude conductivity is given by $\sigma =\frac{ne^2\tau}{m_b}$. From the values of measured conductivity and band mass $m_b$, the relaxation time can be estimated\cite{am}. It turns out to be $30~femtoseconds$. Thus, the decoherence time scale $\tau_D = 10\tau_R$ is $300~femtoseconds$. We need to compare this timescale with the time taken by an electron in hoping from one atom to its nearby atom (the hoping timescale).  The bandwidth of gold is $10~eV$. From tight-binding model, the tunneling matrix element ($t$) is roughly of the order of two electron volts\cite{am}. Thus the hoping time scale is $\frac{\hbar}{2~eV}\sim 0.5~femtoseconds$! This is much smaller than the decoherence time scale ($300~ femtoseconds$). Thus, as an electron hop from an atom to its nearby atom it suffers very small (negligible) amount of decoherence. This is in direct contradiction to stochastic model of Ovchinnikov and Erikhman where stochasticity emerges even in the near neighbor hoping.  On the other hand, hopings over 1000 lattice sites would take roughly $500~femtoseconds$ which is greater than $300~femtoseconds$ (the decoherence time scale from the master equation approach). Therefore, our estimates from the lattice correlation length are roughly in agreement with the estimates from the master equation approach. 

\section{Conclusion}
In conclusion, we are now in a position to answer the question posed: up to what distances electrons in a periodic potential can be considered as constituting an effective closed quantum system? Surely, as electrons hop from atom to atom their motion is fully coherent and decoherence can be safely be neglected. However, over a larger length scale  their motion is no more coherent and can be considered as an open quantum system. Our results from the revision and extension of the Ovchinnikov and Erikhman program are roughly in agreement with those from the master equation approach.

\section{Acknowledgements} 
Author is thankful to Paramita Dutta and Saptarshi Mandal for carefully reading the draft and suggesting corrections and useful comments. 
\begin{wrapfigure}{r}{0.2\textwidth}
  \vspace{-20pt}
  \begin{center}
    \includegraphics[width=0.15\textwidth]{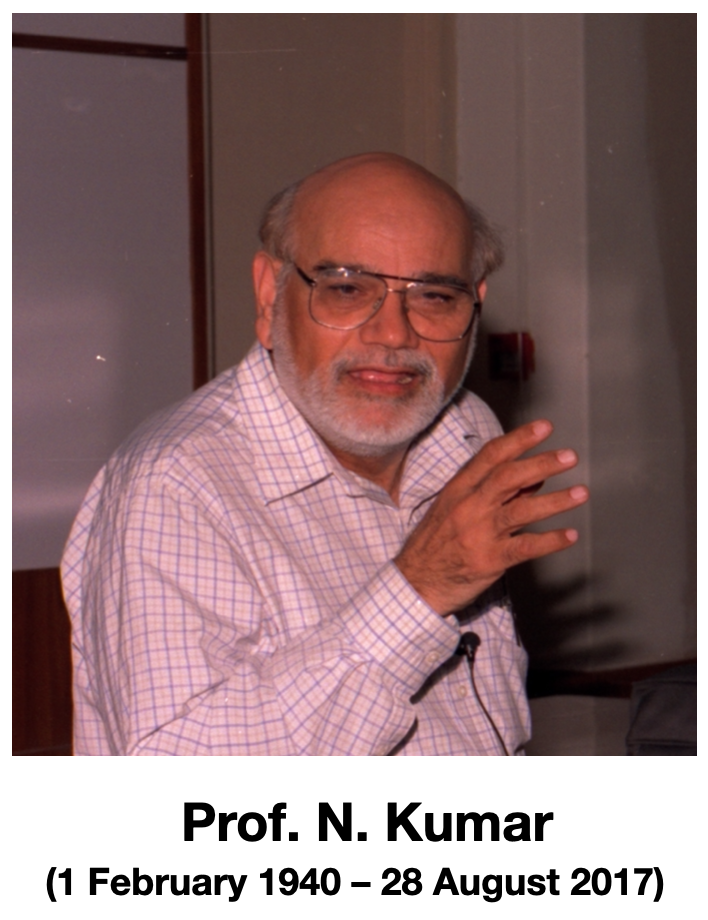}
  \end{center}
  \vspace{-15pt}
  \vspace{-10pt}
\end{wrapfigure}
Author dedicates this manuscript to the loving memory of Prof. N. Kumar (1st Feb 1940 -- 28th Aug. 2017).  Today, 28th August, is his 8th death anniversary. The topic of decoherence was particularly close to his heart.



\end{document}